\pdfoutput=1
\newif\ifFull
\Fullfalse
\documentclass[letterpaper, 11 pt, onecolumn, conference]{ieeeconf}  
\IEEEoverridecommandlockouts                              

\usepackage{microtype}
\usepackage{graphicx}
\usepackage{subfigure}
\usepackage{booktabs} 
\usepackage{amsmath} 
\usepackage{amssymb}  
\newcommand{\E}{{\bf E}}
\newtheorem{theorem}{Theorem}
\newtheorem{definition}[theorem]{Definition}

\usepackage{hyperref}

\begin{document}

\title{\LARGE Scheduling with Predictions and the Price of Misprediction}
\date{}




\author{Michael Mitzenmacher{$^*$} \thanks{{$^*$}School of Engineering and Applied Sciences, Harvard University.  {\texttt{michaelm@eecs.harvard.edu}}.  This work was supported in part by NSF grants CCF-1563710 and CCF-1535795.}}

\maketitle




\vspace{0.25in}

\begin{abstract}
In many traditional job scheduling settings, it is assumed that one
knows the time it will take for a job to complete service.  In such
cases, strategies such as shortest job first can be used to
improve performance in terms of measures such as the average time a
job waits in the system.  We consider the setting where the service
time is not known, but is predicted by for example a machine learning
algorithm.  Our main result is the derivation, under natural
assumptions, of formulae for the performance of several strategies for
queueing systems that use predictions for service times in order to
schedule jobs.  As part of our analysis, we suggest the framework of
the ``price of misprediction,'' which offers a measure of the cost of
using predicted information.  
\end{abstract}

\section{Introduction}

While machine learning research seems to be growing at an exponential
rate, there seems to be surprisingly little overlap with
``traditional'' algorithms and data structures and their analysis.
Here we attempt to bridge this gap for the area of job scheduling,
providing a general framework that may prove useful for additional problems.  

While we begin with settings with a finite number of jobs in order to
provide insight into our approach, our main results are in the area of
queueing systems.  In this setting, we assume there is some algorithm
(such as a neural network or other machine learning algorithm) that
predicts the job time\footnote{We use the terms job time, service time, and
processing time interchangeably; historically these different terms have all
been used.} upon entry; we model this predictor via a
density function $g(x,y)$, so that $g(x,y)$ is the probability density
for a job having actual service time $x$ and predicted service time
$y$. We emphasize that only the predicted service time is known to the
system on the job's arrival, and we need not assume that the joint
distribution is known in order to perform the scheduling. Rather, we
use the joint distribution to derive equations for queue performance.

In standard queueing theory, under standard assumptions such as
Poisson arrivals, and independent service times, one can derive
formulae for the behavior of many natural scheduling strategies,
including {\em shortest job first} (SJF) and {\em shortest remaining
  processing time} (SRPT), which both minimize the average time a job
spends in the system.  (Shortest job first assumes no preemption;
shortest remaining processing time allows preemption.)  In the setting
where job times are predicted, we refer to the corresponding natural
strategies as {\em shortest predicted job first} (SPJF) and {\em
  shortest predicted remanining processing time} (SPRPT).  Our main
result is to derive formulae for the expected time a job spends in the
system for such strategies;  the formulae can be computed in
terms of the density function $g(x,y)$. We further provide some empirical
evidence from simulations that even weak predictions can yield very
good performance.

More generally, we consider the cost of using predictions in place
of accurate job service times, and introduce the concept of the 
{\em price of misprediction} to describe this cost.  Our results
provide the price of misprediction for these basic strategies.

We emphasize that our goal here is {\em not} to develop specific prediction
methods, and we do not do so in this work.  Rather, our goal is to show that given
a real or hypothetical prediction system matching our assumptions, we
can develop equations for its performance.  This general framework
may apply to a variety of machine learning methods.  In this way, we aim to
extend traditional queueing theoretic formulations to the general
setting where machine learning prediction systems are available.  Just
as queueing theoretic models and results have historically guided many real-world
systems (see, e.g., \cite{harchol2013performance,kleinrock1975queueing,kleinrock1976queueing}
for background), our motivation stems from the idea that extending such models and results to setting with predictions
will enhance the use of machine learning prediction in real-world systems that use queues.
Indeed, we further hope that our approach may prove useful for the analysis of other 
traditional algorithms and data structures.  

\section{Related Work}

While traditional algorithmic analysis focuses on worst-case algorithm
behavior, there is a growing movement to develop frameworks that go
beyond worst-case analysis \cite{roughgarden2018beyond}.  While such frameworks have existed in the
past, most notably via probabilistic analysis (e.g.,
\cite{DBLP:books/daglib/0012859}), semi-random models (e.g.,
\cite{blum1995coloring,feige2000finding}), and smoothed analysis
\cite{spielman2004smoothed}, one natural approach that has received
little attention is the use of machine-learning-based approaches to
provide predictions to algorithms, with the goal of realizing provable
performance guarantees.  (The idea of using machine learning to give
hints as to which {\em heuristic} algorithm to employ has been considered in
meta-heuristics for several large-scale problems, most notably for
satisfiability \cite{xu2008satzilla}; this is a distinct line of
work.)

Notable recent work with this theme is that of Lykouris and
Vassilvitskii \cite{DBLP:conf/icml/LykourisV18}, who show how to use
prediction advice from machine learning algorithms to improve online
algorithms for caching in a way that provides provable performance
guarantees, using the framework of competitive analysis.  A series of
recent papers consider the setting of optimization with noise, such as
in settings when sampling data in order to obtain values used in an
optimization algorithm for submodular functions
\cite{DBLP:conf/nips/BalkanskiRS16,DBLP:conf/colt/HassidimS17,
  DBLP:conf/stoc/BalkanskiRS17,DBLP:conf/colt/BalkanskiS17,DBLP:conf/icml/BalkanskiS18,DBLP:conf/icml/RosenfeldBGS18}.
Other recent works analyze the performance of learned Bloom filter
structures \cite{TCFLIS,mitzenmacher2018model}, a variation on Bloom
filters \cite{bloom1970space} that make use of machine learning
algorithms that predict whether an element is in a given fixed set as
a subfilter structure, and heavy hitter algorithms that use
predictions \cite{hsu2018learning}.
One prior work in this vein has specifically
looked at scheduling with predictions in the setting of a fixed collection of
jobs, and considered variants of shortest predicted processing time that 
yield good performance in terms of the competitive ratio, with the performance
depending on the accuracy of the predictions \cite{purohit2018improving}. 

In scheduling, some works have looked at the effects of using
imprecise information, usually for load balancing in multiple queue
settings.  For example, Mitzenmacher considers using old load
information to place jobs (in the context of the power of two choices)
\cite{DBLP:journals/tpds/Mitzenmacher00}. A strategy called TAGS
studies an approach to utilizing multiple queues when no information
exists about the service time; jobs that run more than some threshold
in the first queue are cancelled and passed to the second queue, and
so on \cite{DBLP:journals/jacm/Harchol-Balter02}.  For single queues,
recent work by Scully and Harchol-Balter have considered 
scheduling policies that are based on the amount of service received, where the
scheduler only knows the service received approximately, subject to adversarial noise, and the goal is to develop robust policies \cite{scully2018}.  Our
work differs from these past works in providing a model specifically geared
toward studying performance with machine-learning based predictions, and 
corresponding analyses.

Finally, we note that our policies appear to fit within the more general framework of SOAP policies presented by Scully et~al. \cite{scully2018also} This provides an alternative approach to analyzing the policies studied here.  We present derivations here based on the original analyses of SJF and SRPT, as we feel they are more instructive and straightforward.  

\section{Price of Misprediction}

\subsection{A Simple Example}
To demonstrate our framework, we start with a simple example.  Consider a
collection of $n$ jobs $j_1,\ldots,j_n$, each of one of two types,
short or long.  Short jobs require time $s$ to process and long jobs
require time $\ell$ to process, with $s < \ell$.  Jobs are to be ordered and then
processed sequentially.  When the job times are known, shortest job
first is known to minimize the average waiting time over all jobs.
(Here and throughout waiting time is the time spent in the system
before starting being served.)
If there are $n_s$ short
jobs and $n_\ell$ long jobs the average waiting time is
\begin{align*}
\frac{1}{n} \left ( n_s \frac{n_s -1}{2} s + n_\ell \frac{n_\ell -1}{2} \ell + n_\ell n_s s \right).
\end{align*}
That is, on average each of the $n_s$ jobs waits for half of the remaining $n_s-1$ short jobs, and similarly for the long jobs, and the long jobs further have to wait for all of the short jobs.  

We note that asymptotically if we drop lower order terms this is approximately
\begin{align*}
\frac{n_s^2 s + n_\ell^2 \ell + 2n_s n_\ell s}{2n}.
\end{align*}
For simplicity, we will generally work with asymptotic expressions throughout.

If one has no information about the type of all of the jobs, the optimal policy (under the assumption that an adversary can present the jobs in a worst-case order) is to randomize the order of the jobs.  In this case, using linearity of expectations, we can find the overall expected waiting time 
by finding the expected waiting time of each job.  A simple calculation shows that if there are $n_s$ short jobs and $n_\ell$ long jobs the expected waiting time is
\begin{align*}
\frac{1}{n} \left ( n_s \left( \frac{n_s -1}{2} s + \frac{n_\ell}{2} \ell \right )
+ n_\ell \left( \frac{n_s}{2} s + \frac{n_\ell -1}{2} \ell \right ) \right ).
\end{align*}
Asymptotically, this is approximately
\begin{align*}
\frac{n_s^2 s + n_\ell^2 \ell + n_s n_\ell (s+\ell)}{2n}.
\end{align*}

Finally, in the prediction setting, when a job's type is predicted 
we assume short jobs are misclassified as
long jobs with some probability $p$ and long jobs are misclassified as
short jobs with some probability $q$.  We consider the policy of 
{\em shortest-predicted-job-first}; that is, we apply
shortest-job-first based on the predictions.  Our analysis requires multiple cases,
as we must consider the expected waiting time for a job
conditioned on whether it was classified correctly or incorrectly. 
Considering all of these four cases leads to the following
expression 
for the expected waiting time when mispredictions occur:

{\scriptsize
  \setlength{\abovedisplayskip}{6pt}
  \setlength{\belowdisplayskip}{\abovedisplayskip}
  \setlength{\abovedisplayshortskip}{0pt}
  \setlength{\belowdisplayshortskip}{3pt}
\begin{align*}
& \frac{1}{n} \bigg(  (1-p) n_s \left( \frac{(1-p)(n_s -1)}{2} s + \frac{q n_\ell}{2} \ell \right) \\
& + pn_s \left( (1-p)(n_s -1) s + \frac{p(n_s -1)}{2} s
+ \frac{(1-q) n_\ell}{2}\ell + q n_\ell\ell \right ) \\
& + 
(1-q) n_\ell \left( \frac{(1-q)(n_\ell -1)}{2} \ell + q (n_\ell-1) \ell 
 + \frac{pn_s}{2} s + (1-p) n_s s \right) \\
&  +
  q n_\ell \left( \frac{q(n_\ell-1)}{2} \ell 
  + \frac{(1-p)n_s}{2} s \right) \bigg).
 \end{align*}
}%

Asymptotically, this is approximately
\begin{align*}
\frac{n_s^2 s + n_\ell^2 \ell + n_s n_\ell ((2-(p+q))s+(p+q)\ell)}{2n}.
\end{align*}
This differs from the optimal (asymptotic) expression additively by
$\frac{(p+q)(\ell-s)}{2n}$;  it depends specifically on the ``total error''
$p+q$.

Following the standard terminology\footnote{The terms ``price of anarchy'' \cite{koutsoupias1999worst}
and ``price of stability'' \cite{anshelevich2008price}
are commonly used in game theoretic situations, and in particular in job scheduling,
when multiple players act in their own self interest instead of cooperating.  One could
also view this as a multiplicative form of regret, but we think this terminology is
more general and potentially helpful.}, we might refer to the ratio
between the expected waiting time with imperfect information and the expected
waiting time with perfect information as the {\em price of misprediction}.
We propose the following definition:
\begin{definition}
Let $M_A(Q;I)$ be the value of some measure (such as the expected
waiting time) for a system $Q$ given information $I$ about the system
using algorithm $A$, and let $M_A(Q;P)$ be the value of that metric
using predicted information $P$ in place of $I$ when using algorithm $A$.
Then the price of misprediction is defined as $M_A(Q;I)/M_A(Q;P)$.
\end{definition}

In the example of short and long jobs above, the asymptotic price of 
misprediction for the waiting time is the ratio $R$ given by:
\begin{align*}
R = \frac{n_s^2 s + n_\ell^2 \ell + n_s n_\ell ((2-(p+q))s+(p+q)\ell)} {n_s^2 s + n_\ell^2 \ell + n_s n_\ell (s+\ell)}
\end{align*}
We can find where the ratio is maximized by considering $n_s = \gamma n$ for a constant $\gamma$.
Some algebraic work yields that:
\begin{align*}
R \leq 1 + \frac{(p+q)(\sqrt{l/s}-1)}{2},
\end{align*}
giving a bound on the price of misprediction.  Note that the setting with no information,
or a random ordering, is equivalent to the case $p=q=1/2$, in which case $R$ is bounded
by $1 + (\sqrt{l/s}-1)/2$.  

\subsection{General Predictions}

More generally, we can consider a setting where each job can be described as
an independent random variable, where the random variable for a job is given
by a density distribution $g(x,y)$;  that is, $g(x,y)$ is the density function 
for a job, where a job has service time $x$ and predicted service time $y$.
We assume that $g(x,y)$ is ``well-behaved'' throughout this work, so that it is continuous and all
necessary derivatives exist.  (The analysis can be readily modified to handle point masses or other discontinuities in the distribution.)

It is convenient to let
$f_s(x) = \int_{y=0}^\infty g(x,y) \, dy$ be the density
function for the service time, and
$f_p(y) = \int_{x=0}^\infty g(x,y) \, dx$ be the density
function for the predicted service time.
If there are $n$ total jobs, the expected waiting time for a job using shortest job first given full information is given by
\begin{align*}
(n-1) \int_{x=0}^\infty f_s(x) \left (\int_{z=0}^x z f_s(z) \,dz \right ) dx, 
\end{align*}
while the expected waiting time for a job using predicted information using shortest predicted job first is given by
\begin{align*}
(n-1) \int_{y=0}^\infty f_p(y) \left (\int_{x=0}^\infty \int_{z=0}^y x g(x,z) \, dz \, dx \right ) dy. 
\end{align*}
In words, in the full information case, to compute the expected
waiting time for a job, given its service time, we can determine the
probability each other job has a smaller service time and the
conditional value of that smaller service time to compute the waiting
time.  In the predicted information case, to compute the expected
waiting time for a job given its predicted service time, we must
determine the probability each other job has a smaller predicted
service time and the conditional value of the actual service time of
a job given that it has a smaller predicted service time to
compute the waiting time.


Note that the factors of $n-1$ are cancelled in the
ratio of the expected waiting times, and the function $g$ suffices to determine to price of misprediction.  
As an example motivated by the prevalence of exponential distributions in queueing theory, suppose that service times are exponentially
distributed with mean 1, and a job with service time $x$ has a
prediction that is distributed according to an exponential
distribution with mean $x$.  Then $g(x,y) = e^{-x-y/x}/x$, $f_s(x) =
e^{-x}$, and $f_p(y) = \int_{x=0}^\infty \frac{e^{-x-y/x}}{x} \, dx$.
We note $f_p(y)$ does not appear to have a simple closed form, though
it is expressible in terms of Bessel functions.\footnote{This was
determined using the integration features of Mathematica 11.3.}
We note numerical calculations from these integrals readily reveal
that the price of misprediction appears to be $4/3$, and it can be shown to be exactly $4/3$
using the integration features of Mathematica. We can prove it is exactly equal to
$4/3$ through a subtle argument
allowing us to evaluate the corresponding integral for 
the expected waiting time for a job using predicted information;  this
argument is given in the appendix.   In comparison, using no information and just scheduling in sequential order the expected waiting time is a factor of 2 worse than
when using full information.  This example, while not meant to match a real-world example,provides the right high-level intuition, in that it shows that even a weak predictor
can yield significant improvements.  Indeed, this is natural; for a predictor to work well in this setting, it simply has to order most of the jobs correctly in the queue.  

The key here is that the price of misprediction can be computed (at
least numerically) given the density distribution $g$.  In practice,
one might use this framework to determine the benefit of using a
predictor; for example, one might seek to trade off the reduction in
total waiting time with the cost of developing or using better prediction
methods.  While $g$ may not be known exactly, we expect in practice
good approximations for $g$ can be determined empirically, which in
turn will allow a good approximation for the price of misprediction or
related quantities.

We note that, for suitably good prediction schemes, ordering by predicted service
time should naturally correspond to ordering by the expected service time.
That is, denote a job by $(X,Y)$, where $X$ is a random variable representing
the true service time and $Y$ is a random variable representing the predicted service time.
Then suppose the density distribution $g$ satisfies for any $y_1, y_2$ with 
$y_1 < y_2$ the natural inequality
\begin{align*}
\E[X~|~Y=y_1] < \E[X~|~Y=y_2].
\end{align*}
Then ordering by predicted service times yield an ordering according to expected
service times, and ordering by expected service times is known to be optimal for 
minimizing the expected waiting time.

We now extend these ideas to queueing theoretical models.  

\section{Single Queue Models}

In this section, we present results providing formulae for
prediction-based variants of shortest job first and shortest remaining
processing time for single queue systems, which yield expressions
for the price of misprediction.  We briefly review the appropriate 
analysis methods for standard queues, starting with jobs with priorities,
and then extend them prediction setting.

\subsection{Priority-based Systems}

Consider a queueing system with $k$ types of jobs, $t_1,\ldots,t_k$.
We assume Poisson arrivals, and that the arrival rate for type $t_i$
is $\lambda_i$, with $\sum_{i=1}^k \lambda_i = \lambda$.  A natural
setting is that the $i$th type of job has service time $q_i$, with
$q_1 < q_2 < \ldots < q_k$.  In this case, with complete information
about the service times, the shortest job first (SJF) strategy
(without preemption) corresponds to a priority-based strategy, with
the types corresponding to priorities; $t_1$ has the highest
priority, and so on.  More generally, the $i$th type of job may have a service
time distribution, rather than a fixed service time.  We describe this
more general case, where job $t_i$ has service distribution $S_i$;
here the natural setting is $\E[S_1] < \E[S_2] < \ldots < \E[S_k]$,
and if the types are prioritized by expected service time, 
the strategy is expected shortest job first (ESJF).


We describe some standard formula for priority systems, following the 
framework of \cite{harchol2013performance}.
Let $\rho_i = \lambda_i \E[S_i]$;  this represents the load on the system
from jobs of type $i$.  Further, let 
$\rho = \sum_{i=1}^k \rho_i$, 
and $W(i)$ be the distribution of the waiting time in the queue for jobs of type $i$
in equilibrium.
Also, let $S$ be service time distribution of an incoming job,
so $\E[S] = \sum_{i =1}^k \lambda_i \E[S_i] / \lambda$, 
and $\rho = \lambda \E[S]$.  
Then the following is known (see Equation (31.1) of \cite{harchol2013performance}):
\begin{align*}
\E[W(i)] = \frac{\rho \E[S^2]}{2\E[S] \left ( 1 - \sum_{j =1}^i \rho_j \right )  \left ( 1 - \sum_{j =1}^{i-1} \rho_j \right )}.
\end{align*}

We now consider a system where types are not known but are predicted,
for example according to a machine learning algorithm.  For convenience, going
forward, we refer to the {\em true type} of a job for its type, and
refer to the machine prediction for a job as the {\em predicted type}
where appropriate.  We may represent the machine learning algorithm by
a matrix $M$ where $m_{ij}$ is the probability that a job of true type $i$
has predicted type $j$;  here we are assuming that each job labelling can
be treated as independent.  In this case, let $\lambda'_i$ be the arrival
rate of jobs with predicted type $i$.  Then
\begin{align*}
\lambda'_i = \sum_{j=1}^k \lambda_j m_{ji}.  
\end{align*}
Correspondingly, the distribution of service times for jobs having predicted
type $i$ is that the job has service time given by $S_\ell$ with probability 
\begin{align*}
\lambda_\ell m_{\ell i}/ \sum_{j=1}^k \lambda_j m_{ji}.  
\end{align*}
If we use $S'_i$ to represent the distribution of service times
for jobs having predicted type $i$, then 
\begin{align*}
\E[S'_i] = \frac{\sum_{j = 1}^k \lambda_j \E[S_j] m_{j i}}
{\sum_{j=1}^k \lambda_j m_{j i}}.    
\end{align*}
Of course the expected service time $S'$ over all jobs is
\begin{align*}
\E[S'] = \sum_{i = 1}^k \lambda_i \E[S_i] = \E[S].  
\end{align*}

Assuming we prioritize jobs now according to their predicted type, 
we may again use the standard formula for priority systems.
We derive the corresponding result.  First, let
\begin{align*}
\rho'_i = \lambda'_i \E[S_i] = \sum_{j = 1}^k \lambda_j \E[S_j] m_{j i}.
\end{align*}
Also let $W'(i)$ be the distribution of the waiting time in the queue for jobs of predicted
type $i$ in equilibrium.  
Then 
\begin{align*}
\E[W'(i)] & = \frac{\rho' \E[(S')^2]}{2\E[(S')] \left ( 1 - \sum_{j =1}^i \rho'_j \right )  \left ( 1 - \sum_{j =1}^{i-1} \rho'_j \right )} \\
 & = \frac{\rho \E[S^2]}{2\E[S] \left ( 1 - \sum_{j =1}^i \rho'_j \right )  \left ( 1 - \sum_{j =1}^{i-1} \rho'_j \right )}.
\end{align*}

Hence, by summing over all possible types, we can see that the price of misprediction for the expected waiting time corresponds
to the following expression:
\begin{align*}
\frac {\sum_{i=1}^k \lambda'_i \left ( \left ( 1 - \sum_{j =1}^i \rho'_j \right )  \left ( 1 - \sum_{j =1}^{i-1} \rho'_j \right ) \right )^{-1}}
{\sum_{i=1}^k \lambda_i \left ( \left ( 1 - \sum_{j =1}^i \rho_j \right )  \left ( 1 - \sum_{j =1}^{i-1} \rho_j \right ) \right )^{-1}}.
\end{align*}

\subsection{Shortest Job First}

We now show that the performance of shortest predicted job first,
which we denote as SPJF, can be readily expressed as a limiting case of
the priority analysis, similarly to how shortest job first is the limiting
case of a priority queue based on service time.  (Here we roughly follow
the methodology of Section 31.3 of \cite{harchol2013performance}.)   To start, we recall
the formula for shortest job first; this is easily obtained as the
limit of the priority system setting, where there are an infinitely
many possible ``priorities'', and the priority corresponds to the
service time.  Here again let $S$ be the service distribution of an
incoming job.  Further, let $f_s(x)$ be the corresponding density
function, $\rho_x = \lambda \int_{t=0}^x tf_s(t) dt$, and
$\rho = \lambda \int_{t=0}^\infty tf_s(t) dt$.  We 
consider $W(x)$, the time spent waiting in the queue (not being served)
for jobs with service time
$x$ in equilibrium.  Then for standard shortest job first without preemption, where
we know the exact service times without prediction, it is known that 
\begin{align*}
\E[W(x)] = \frac{\rho \E[S^2]}{2\E[S] \left ( 1 - \rho_x \right )^2}.
\end{align*}
The overall expected time waiting in a queue, which
we denote by $\E[W]$ where $W$ is the waiting time in queue of an incoming job, is then simply
\begin{align*}
\E[W] = \int_{x=0}^\infty f(x) \E[W(x)] \, dx.
\end{align*} 

We now generalize this to SPJF.
For a non-preemptive queue that uses a service time
estimate, if $g(x,y)$ is the joint distribution that a
job has service time $x$ and predicted service time $y$,
we again let 
$f_s(x) = \int_{y=0}^\infty g(x,y) \, dy$ be the density
function for the service time, and
$f_p(y) = \int_{x=0}^\infty g(x,y) \, dx$ be the density
function for the predicted service time.
$\rho'_y = \lambda \int_{t=0}^y \int_{x=0}^\infty  x g(x,t) \, dx \, dt$  to be the load 
on the system associated with jobs of predicted service time up to $y$.  
With the assumption that each job's service time characteristics
are independently determined according to $g(x,y)$, if we let
$W'(y)$ be the distribution of time spent waiting in the queue for a job with predicted service time $y$
in equilibrium,
then 
\begin{align*}
\E[W'(y)] = \frac{\rho \E[S^2]}{2\E[S] \left ( 1 - \rho'_y \right )^2},
\end{align*}
where $W'(y)$ is the distribution of time in the queue for jobs 
with service time $y$.  Integrating over service time
or predicted services time gives us that the  
price of misprediction is given by:
\begin{align*}
\frac{\int_{y=0}^\infty \frac{f_p(y)}{(1-\rho'_y)^2} \, dy}
{\int_{x=0}^\infty \frac{f_s(x)}{(1-\rho_x)^2} \, dx} .
\end{align*}
Let us again consider the example of service times that are
exponentially distributed with mean 1, where a job with service time
$x$ has a prediction that is distributed according to an exponential
distribution with mean $x$.  Then the price of misprediction can be expressed as
\begin{align*}
\frac{\int_{y=0}^\infty \frac{\int_{x=0}^\infty \frac{e^{-x-y/x}}{x} \, dx}{(1-\lambda \int_{t=0}^y \int_{x=0}^\infty e^{-x-y/x} \, dx \, dt)^2} dy}
{\int_{x=0}^\infty \frac{e^{-x}}{(1-\lambda(1-(x+1)e^{-x}))^2} \, dx}.
\end{align*}
While there does not appear to be a simple closed form for this expression,
it can be readily evaluated numerically for a given $\lambda$.    

We note that a similar analysis can be used for preemptive 
shortest predicted job first (PSPJF), where a job may be preempted by another
job that has an originally shorter predicted time (note that the time
a job has been serviced is not considered).  This is because
preemptive shortest job first (PSJF) can be represented as the
limit of a preemptive priority-based system (as in Section 32.3 of \cite{harchol2013performance}), leading to a similar
analysis.  (We provide the analysis in the appendix.)  Also, we can consider
variations where the machine learning algorithm returns a distribution
(described by a small number of parameters) as a prediction; for
example, the prediction might be the service time is exponential with
mean $\gamma$.  In this case, we can use the shortest predicted 
expected processing time (SPEPT), which reduces readily to the analysis of
shortest predicted job first.

\subsection{Shortest Remaining Processing Time}

A more challenging variation involves extending the shortest remaining
processing time (SRPT) policy to predictions.  With complete
information, SRPT maintains the remaining processing time for each
job, and the job being processed can be preempted by an incoming job
with service time smaller than the remaining processing time.  To
generalize to the prediction setting, we follow the framework of
Schrage and Miller \cite{schrage1966queue}, who presented an analysis of SRPT.  (See also
\cite{gautam2012analysis} for a similar derivation, or \cite{harchol2013performance} for
an alternative.)  Because the system is preemptive,
it makes sense to consider the total time in the system, rather than
the waiting time (as jobs may have further waits after they start
being served).  Because of the complexity of the expressions, we do
not have a clean form for the price of misprediction, but they can be
found from the derived formulae.

We again use $g(x,y)$ for the joint distribution that a job has
service time $x$ and predicted service time $y$, and let 
$f_s(x) = \int_{y=0}^\infty g(x,y) \, dy$, 
$f_p(y)
= \int_{x=0}^\infty g(x,y) \, dx$, 
$\rho_x = \lambda \int_{t=0}^x tf_s(t) \, dt$,
and $\rho'_y
= \lambda \int_{t=0}^y \int_{x=0}^\infty g(x,t) x \, dx \, dt$.
The expected time in the system in equilibrium for a job can be expressed as the
sum of its residence time (time in the system once it has started receiving service)
and it waiting time (time spent waiting before being served).
For SRPT, a job with service time $x$ has mean residence time
\begin{align*}
\int_{t=0}^x \frac{dt}{1-\rho_t};
\end{align*}
one can think of this as the remaining service times drop from $t$ to 0,
at any instant there is a possible addition (given by the $1/(1-\rho_t)$
factor) due to preemptions. 

The corresponding mean residence time for a job of
service time $x$ and predicted service time $y$ under SPRPT is
\begin{align*}
\int_{t=0}^x \frac{dt}{1-\rho'_{(y-t)^+}}.
\end{align*}
That is, here the predicted remaining processing time drops from $y$ to $(y-t)^+
= \max(y-t,0)$;  it is possible the predicted remaining processing time
is 0, but the job continues to require service, in which case we leave
its predicted remaining processing time at 0, and it cannot be preempted.  
It follows that the mean residence time $\E[R(y)]$ for a job of predicted service time $y$
is
\begin{align*}
\E[R(y)] = \int_{x=0}^\infty \frac{g(x,y)}{f_p(y)} \int_{t=0}^x \frac{dt}{1-\rho'_{(y-t)^+}} \, dx.
\end{align*}

We now compute the waiting time, which is more difficult.
The steady-state probability that an arriving job finds
the server working on a job whose remaining predicted 
processing time is less than $q$ is given by
\begin{align*}
b(q) = \rho'_q + \lambda \int_{t=q}^\infty \int_{x=0}^\infty g(x,t) (x-(t-q))^+ \, dx \, dt.  
\end{align*}
The first term comes from arrivals with predicted service time less than $q$;  
the second term comes from jobs that start with predicted service time greater than
$q$, but later their remaining predicted service time falls below $q$.  

If $Y(q)$ is the length of a busy period where all jobs processed have
predicted remaining processing times less than $q$, then the waiting
time $W(q)$ for a job of predicted service time $q$ is given by:
\begin{align*}
\E[W(q)] = b(q) \frac{\E[Y(q)^2]}{2\E[Y(q)]}.  
\end{align*}
To find the first two moments of $Y(q)$, we use the fact that the length of the busy
period $Y(q)$ has the same distribution as the busy period for a
first-come, first-served server where the job that initiates the busy
period has processing time according to some distribution $Z(q)$,
where additional jobs have a processing time distribution $X(q)$, and
the arrivals are Poisson with rate $\lambda F_p(q)$, for $F_p(q) =
\int_{x=0}^q f_p(x) dx$.  We require the first two moments of $Z(q)$ and
$X(q)$.

The moments for $X(q)$ are fairly straightforward, as $X(q)$ corresponds
to the processing time of a job with predicted processing time at most
$q$:
\begin{align*}
\E[X(q)] = \frac{1}{F_p(q)} \int_{t=0}^q \int_{x=0}^\infty g(x,t) x \, dx \, dt,
\end{align*}
and 
\begin{align*}
\E[X(q)^2] = \frac{1}{F_p(q)} \int_{t=0}^q \int_{x=0}^\infty g(x,t) x^2 \, dx \, dt.
\end{align*}

To determine the first two moments of $Z(q)$, we note that there are two
ways a job can start the corresponding busy period.  It either arrives
when a busy period is not in progress and has predicted processing time at most $q$, or it is a job with predicted processing time greater than $q$ (which starts a busy period when the predicted processing time reaches $q$).  
Note that in the second case, if the predicted processing time $t$ is
greater than $q$, but the actual processing time $x$ is such that 
$x < t - q$, then the job cannot start a busy period, as the job will finish
before the remaining predicted processing time reaches $q$.  (Ideally, such
situations should not occur with a suitably good predictor, but it must be
taken into account.)  Hence the probability a job initiates a corresponding
busy period is 
\begin{align*}
d(q) = (1-b(q))F_p(q) + \int_{t=q}^\infty \int_{x=t-q}^\infty g(x,t) \, dx \, dt.
\end{align*}
If we let (for typesetting reasons)
\begin{align*}
a_1(q) & = (1-b(q)) \int_{t=0}^q \int_{x=0}^\infty g(x,t) x \, dx \, dt \\
& + \int_{t=q}^\infty \int_{x=t-q}^\infty g(x,t) (x-(t-q)) \, dx \, dt
\end{align*}
and 
\begin{align*}
a_2(q) & = (1-b(q)) \int_{t=0}^q \int_{x=0}^\infty g(x,t) x^2 \, dx \, dt \\
& + \int_{t=q}^\infty \int_{x=t-q}^\infty g(x,t) (x-(t-q))^2 \, dx \, dt
\end{align*}
then 
\begin{align*}
\E[Z(q)] = a_1(q)/d(q),
\end{align*}
and 
\begin{align*}
\E[Z(q)^2] = a_2(q)/d(q).
\end{align*}

We now use the facts (see, e.g., Problem 49 of \cite{gautam2012analysis})
\begin{align*}
\E[Y(q)] = \frac{\E[Z(q)]}{1-\rho'_q}
\end{align*}
and 
\begin{align*}
\E[Y(q)^2] = \frac{\E[Z(q)^2]}{(1-\rho'_q)^2} + \lambda \E[Z(q)] F_p(q) \frac{\E[X(q)^2]}{(1-\rho'_q)^3}.
\end{align*}
This yields
\begin{align*}
\E[W(q)] = b(q) \left (\frac{a_2(q)}{2a_1(q)(1-\rho'_q)} + \lambda F_p(q) \frac{\E[X(q)^2]}{2(1-\rho'_q)^2} \right ).
\end{align*}

The expected time in the system for a job is simply $\int_{y=0}^\infty
f_p(y) \E[W(y)+R(y)] dy$.  From this value (and the corresponding
equations for standard SRPT) one can compute the price of
misprediction for the total expected time in the system.
(Of course, the expected service time can be subtracted if desired.)

\section{Simulation Results}
We present a small number of simulation results to demonstrate that
our equations are accurate and, at least in the cases we have
examined, the price of misprediction is generally small. 
Correspondingly, this implies that even a small amount of predictive power yields
significantly better performance than standard First-In First-Out
(FIFO) queueing.  We focus on high load settings, as under low load
all systems perform well.  We also note that additional simulations we
have performed further substantiate our high-level conclusions.

We first compare simulation results against the results from our
equations; we also provide results for schemes with full information
for comparison.  Our results are for the setting with Poisson arrivals, service
times are exponential with mean 1, and predicted service times are
exponential with mean $x$ when the actual service time is $x$.  
For consistency, we provide the total expected time in the system (waiting
and service). The results of the equations were computed using Mathematica 11.3 and
numerical integration.  The calculations for SPRPT are somewhat
lengthy and can lead to numerical stability issues; we found
integrating up to predicted times of at most 50 gives accurate answers
while being computable in reasonable time, approximately half an hour on a modern laptop. (Predicted service times greater
than 50 are very rare;  they occur with probability less than $5 \cdot 10^{-7}$.)
We did not optimize the calculations and expect this could be improved.  The results of
simulations were from our own implementation of a queue simulator.
The simulations are the results of averaging the average time over 1000 trials, where in each trial
we recorded the time in system of each completed job.  The trials
were each run for $1\,000\,000$ time units, with jobs completing in the
first $100\,000$ time units discarded from the calculations of the averages to
remove bias from starting with an empty system.

\begin{table}
\begin{center}
{\small
\begin{tabular}{|c|c|c|c|c|c|} 
\hline
 & SJF & SJF & SPJF & SPJF & FIFO \\
$\lambda$  & Eqns & Sims & Eqns & Sims & Eqns\\ [0.5ex] 
 \hline\hline
0.5 & 1.7127 & 1.7128 & 1.7948 & 1.7949 & 2.00\\  \hline
0.6 & 1.9625 & 1.9625 & 2.1086 & 2.1087 & 2.50\\  \hline
0.7 & 2.3122 & 2.3121 & 2.5726 & 2.5730 & 3.33\\  \hline
0.8 & 2.8822 & 2.8828 & 3.3758 & 3.3760 & 5.00\\  \hline
0.9 & 4.1969 & 4.1987 & 5.3610 & 5.3609 & 10.00\\  \hline
0.95 & 6.2640 & 6.2701  & 8.6537 & 8.6541 & 20.00\\  \hline
0.98 & 11.2849 & 11.2734 & 16.9502 & 16.9782 & 50.00\\  \hline
0.99 & 18.4507 & 18.4237 & 29.0536 & 29.1162 & 100.00\\  \hline
\end{tabular}
}
\caption{Results from simulations and equations for Shortest Job First (SJF)
and Shortest Predicted Job First (SPJF).}
\label{tab:table1}
\end{center}
\end{table}

Table~\ref{tab:table1} shows both that the results from equations for SPJF match 
very closely to the simulation results, and that the performance is 
not too much worse than when the service times are known.  With regards
to accuracy, the difference is less than 1\%.  With regard to performance,
using predicted times naturally becomes increasingly worse as load grows,
but the difference still shows the benefits of using imperfect information. 
Recall that, with no information, for standard queueing schemes such as FIFO
the expected time in the system is $1/(1-\lambda)$;  for example, $\lambda = 0.99$ leads to an expected time in the system of $100$.  We see that under high loads,
the gains from prediction remain substantial.

Table~\ref{tab:table2} shows similar results for the same simulation
setting using SRPT and SPRPT.  For SPRPT, the results from equations
align a little less closely to the simulation results, but the
difference remains than 1\%.  Given the complexity of the equations,
and the higher variability in the time in system for SPRTP, this is unsurprising.

\begin{table}
\begin{center}
{\small
\begin{tabular}{|c|c|c|c|c|c|} 
\hline
 & SRPT & SRPT & SPRPT & SPRPT & FIFO\\
$\lambda$  & Eqns & Sim & Eqns & Sim & Eqns\\ [0.5ex] 
 \hline\hline
0.5 & 1.4254 & 1.4251 & 1.6531 & 1.6588 & 2.00\\  \hline
0.6 & 1.6041 & 1.6039 & 1.9305 & 1.9397 & 2.50\\  \hline
0.7 & 1.8746 & 1.8757 & 2.3539 & 2.3684 & 3.33\\  \hline
0.8 & 2.3528 & 2.3519 & 3.1168 & 3.1376 & 5.00 \\  \hline
0.9 & 3.5521 & 3.5486 & 5.04808 & 5.0973 & 10.00\\  \hline
0.95 & 5.5410 & 5.5466  & 8.3221 & 8.4075 & 20.00\\  \hline
0.98 & 10.4947 & 10.5003 & 16.6239 & 16.7852 &50.00\\  \hline
0.99 & 17.6269 & 17.6130  & 28.7302 & 28.7847 & 100.00\\  \hline
\end{tabular}
}
\caption{Results from simulations and equations for Shortest Remaining Processing Time (SRPT)
and Shortest Predicted Remaining Processing Time (SPRPT).}
\label{tab:table2}
\end{center}
\end{table}

\begin{figure}
    \centering
    \includegraphics[width=3.3in]{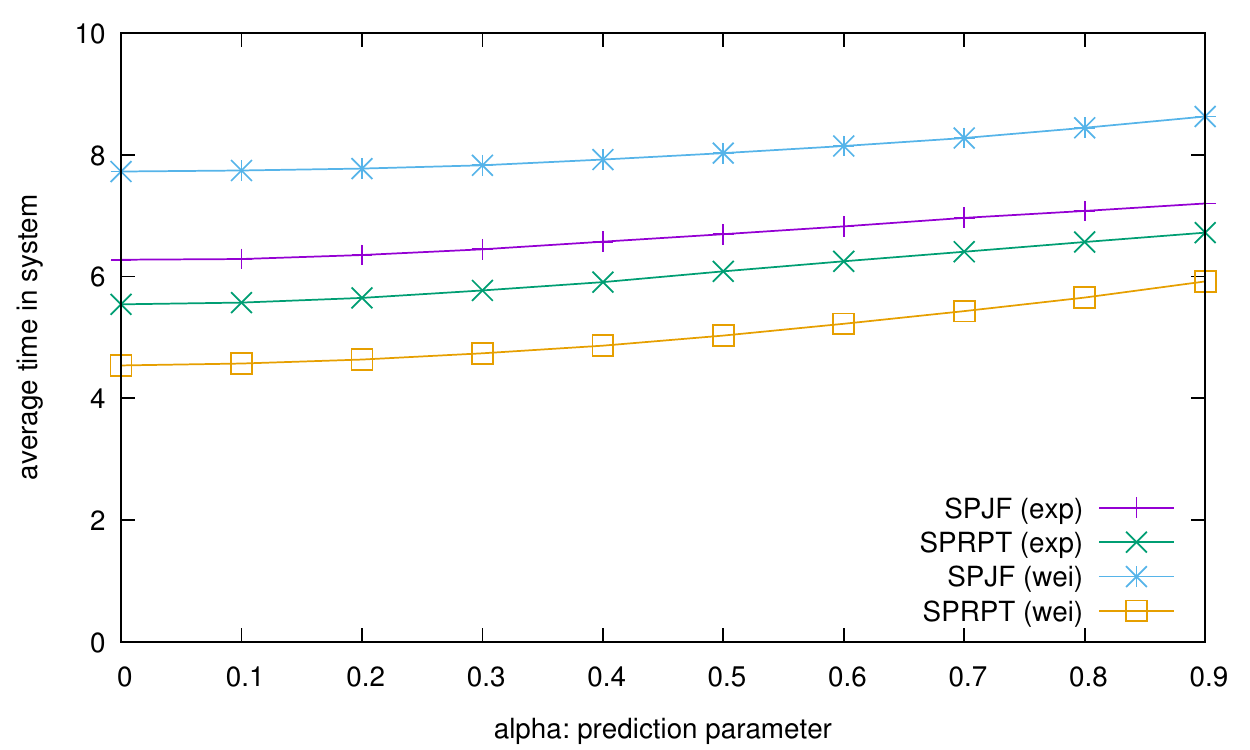}
    \caption{Results from simulations at $\lambda = 0.95$ for exponential and Weibull distributions.
A job with service time $x$ has predicted service time uniform over $[(1-\alpha)x,(1+\alpha)x]$.
Performance degrades gracefully with $\alpha$.  Note $\alpha = 0$ corresponds to the full information
case, as then the predicted service time equals the true service time.}
    \label{fig:fig1}
\end{figure}

Figure~\ref{fig:fig1} provides another example of prediction
performance.  Here we fix $\lambda = 0.95$, and consider two types of
service distributions: exponential with mean 1, and a Weibull
distribution with cumulative distribution $1-e^{-\sqrt{2x}}$.  (The
Weibull distribution is more heavy-tailed, but also has mean 1.)  The
simulations are again the average of the measured time in system,
averaged from results of 1000 trials, in the
same manner as previously.  
Here the predictions depend on a scale
parameter $\alpha$; a job with service time $x$ has a predicted service
time that is uniform over $[(1-\alpha)x,(1+\alpha)x]$.  By varying
$\alpha$, we can see the impact on performance as prediction accuracy
diminishes.  Note that when $\alpha = 0$ the predicted service time
equals the true service time.  In these examples, we observe that
performance degrades gracefully with $\alpha$, a feature we see 
across values of $\lambda$ in other experiments not presented.  The main point here is that even weak
predictors may perform well under SPJF and SPRPT;  as long as they
generally lead jobs to be processed in the right order, they can yield
substantial benefits.  (We note the standard deviation over
trials ranges from 2-4\%, with higher variance for simulations with 
the Weibull distribution.)


\section{Conclusion}

We have demonstrated that the analyses of various single-queue job
scheduling approaches can be generalized to the setting where
predicted service times are used in place of true values, under the assumption
that the predictions can be modeled as joint distribution with a
corresponding density function.  Such analyses can be used to
determine the price of misprediction, or the potential benefits of
better prediction, for such systems.  

In future work, we plan to provide analyses of multiple queue systems
using predicted service times.  Multiple queue systems are quite
common in practice, but can have more highly variable performance
depending on how the workload is divided among queues.  Natural
strategies to consider include the power of two choices \cite{DBLP:journals/tpds/Mitzenmacher01} and 
size interval task assignment (SITA) \cite{harchol1999choosing}.  We expect
analysis of such systems may require additional techniques, but will
show that in this setting also even mildly accurate predictions can
provide significant value.

In the problems considered here, we were able to determine {\em exact}
formulae for performance, based on our probabilistic assumptions.  It
would be interesting to consider more general job scheduling scenarios 
with fewer assumptions, perhaps using
methods more akin to online analysis, as in \cite{DBLP:conf/icml/LykourisV18}.

We believe this work suggests there is great potential in analyzing the large variety of
job scheduling problems, as well as other similar traditional algorithmic
problems, in the context of prediction.


\newpage


\bibliographystyle{plain}

\newpage
\appendices
\section{Proof of $4/3$ price of misprediction for the finite case }

We recall the setting where there are $n$ jobs, service times are exponential 
with mean 1, the predicted service times are exponential with mean $x$ when
the actual service time is $x$, and we seek to determine the expected waiting time.
Since we care only about the expectation, we may consider the expected waiting
time with just a pair of jobs;  linearity yields the price of misinformation
is the same.  

When using the correct service times, the expected waiting time of a job with the shortest job first is
\begin{align*}
\int_{x=0}^\infty e^{-x} \left (\int_{z=0}^x z e^{-z} \, dz \right ) \, dx, 
\end{align*}
which is easily found to evaluate to $1/4$.  
When using predicted service times, the expected waiting time is 
\begin{align*}
\int_{y=0}^\infty f_p(y) \left (\int_{x=0}^\infty \int_{z=0}^y e^{-x-z/x} \, dz \, dx \right ) \, dy, 
\end{align*}
where $f_p(y) = \int_{x=0}^\infty \frac{e^{-x-y/x}}{x} \, dx$.
This does not appear to evaluate to a closed form of simple functions.  However, suppose
we use as our prediction an exponentially distributed service time with mean $1/x$ instead
of $x$.  This effectively reverses the predicted order, but leads to an easier integral
calculation. Since the expected waiting time for a job over both orderings is trivially 1, finding
the expected waiting time for the reverse order suffices.  

For the reversed order, 
\begin{align*}
f_p(y) & = \int_{x=0}^\infty xe^{-x-yx} \, dx \\
       & = \frac{1}{(y+1)^2},
\end{align*}
and the integral becomes 
\begin{align*}
& \int_{y=0}^\infty \frac{1}{(y+1)^2} \left (\int_{x=0}^\infty \int_{z=0}^y x^2e^{-x-xz} \, dz \, dx \right ) \, dy \\
& =\int_{y=0}^\infty \frac{1}{(y+1)^2} \left (\int_{x=0}^\infty (xe^{-x}-xe^{-x-yx}) \, dx \right) \, dy \\ 
& = \int_{y=0}^\infty \frac{1}{(y+1)^2} \left (\int_{x=0}^\infty (xe^{-x}-xe^{-x-yx}) \, dx \right) \, dy \\ 
& = \int_{y=0}^\infty \left (\frac{1}{(y+1)^2} - \frac{1}{(y+1)^4}\right ) \, dy \\
& = 2/3.
\end{align*} 
The expected waiting time where predictions are exponential with mean $x$ is therefore $1/3$,
and the price of misprediction is $4/3$ as claimed. 

\section{Derivation for PSPJF}

We consider the expected time a job spends in the system in equilibrium for
preemptive shortest predicted job first (PSPJF), where a job may be
preempted by another job that has an originally shorter predicted time
(note that the time a job has been serviced is not considered). 
The analysis is similar to both SPJF and SPRPT. 

Here we consider the expected waiting time and the expected residence
time in steady-state.  
We again use $g(x,y)$ for the joint distribution that a job has
service time $x$ and predicted service time $y$, and let 
$f_s(x) = \int_{y=0}^\infty g(x,y) \, dy$, 
$f_p(y)
= \int_{x=0}^\infty g(x,y) \, dx$, 
$F_p(y) = \int_{t=0}^y f_p(y) \, dt$,
and $\rho'_y
= \lambda \int_{t=0}^y \int_{x=0}^\infty g(x,t) x \, dx \, dt$.
As a job will be preempted by another job with smaller predicted
service time, the mean residence time for a job of
service time $x$ and predicted service time $y$ is
\begin{align*}
\frac{x}{1-\rho'_{y}}.
\end{align*}
This is because the residence time with preemptions is the same as the busy
period started by a job of length $x$ and predicted length $y$, 
where the only jobs that need to be considered in the busy period
have predicted length at most $y$.  This leads 
to the additional $1/(1-\rho'_{y})$ factor.

It follows that the mean residence time $\E[R(y)]$ for a job of predicted service time $y$
is
\begin{align*}
\E[R(y)] = \int_{x=0}^\infty \frac{x g(x,y)}{f_p(y)(1-\rho'_{y})} \, dx.
\end{align*}

The expected waiting time for a job with predicted service time $y$ is
the same as for a shortest job first system, except that the job only
waits for jobs of predicted service times as most $y$.  It follows
that 
\begin{align*}
\E[W(y)] = \frac{\lambda \int_{t=0}^y f_p(t)t^2 \, dt}{2 \left ( 1 - \rho'_y \right )^2}.
\end{align*}
Note that here we have simplified the expression, which would originally
have had a factor
\begin{align*}
\rho'_{y} \frac{\left (\int_{t=0}^y t^2 f_p(t) \,dt \right ) /F_p(y)}{\left (\int_{t=0}^y t f_p(t) \, dt \right ) /F_p(y)}.
\end{align*}
The integral expressions are the second and first moments of the expected service
time for a job with predicted service time at most $y$.  
As $\rho'_{y} = \lambda \int_{t=0}^y t f_p(t) \, dt$, the expression for $\E[W(y)]$
follows. 

The expected time in the system for a job is then again simply $\int_{y=0}^\infty
f_p(y) \E[W(y)+R(y)] \, dy$.

\end{document}